\documentclass[aps,prl,longbibliography,twocolumn,superscriptaddress]{revtex4-2}
\usepackage{amsmath,amssymb}
\usepackage{graphicx}
\usepackage{subfigure}
\usepackage{verbatim}
\usepackage{amsfonts}
\usepackage{dcolumn}
\usepackage{bm}
\usepackage{color}
\usepackage[colorlinks,citecolor=blue]{hyperref}
\usepackage{mathrsfs}
\usepackage{float}

\usepackage{multirow}

\newcommand{\beq}{\begin{eqnarray} }
\newcommand{\eeq}{\end{eqnarray} }
\newcommand{\Beq}{\begin{eqnarray*} }
\newcommand{\Eeq}{\end{eqnarray*} }
\newcommand{\Bmat}{\left(\begin{matrix}}
\newcommand{\Emat}{\end{matrix}\right)}

\begin{document}

\title{Pair-density-wave phase of strongly interacting electrons on the triangular lattice: \\ A variational Monte Carlo study}

\author{Jiucai Wang}
\affiliation{School of Physics, Hangzhou Normal University, Hangzhou 311121, China}
\affiliation{Institute for Advanced Study, Tsinghua University, Beijing 100084, China}

\author{Wen Sun}
\affiliation{Institute for Advanced Study, Tsinghua University, Beijing 100084, China}

\author{Hao-Xin Wang}
\affiliation{Institute for Advanced Study, Tsinghua University, Beijing 100084, China}

\author{Zhaoyu Han}
\affiliation{Department of Physics, Stanford University, Stanford, CA 94305, USA}

\author{Steven A. Kivelson}
\email{kivelson@stanford.edu}
\affiliation{Department of Physics, Stanford University, Stanford, CA 94305, USA}

\author{Hong Yao}
\email{yaohong@tsinghua.edu.cn}
\affiliation{Institute for Advanced Study, Tsinghua University, Beijing 100084, China}

\date{\today}

\begin{abstract}
A robust theory of the mechanism of pair density wave (PDW) superconductivity (i.e. where Cooper pairs have nonzero center of mass momentum) remains elusive. Here we explore the triangular lattice $t$-$J$-$V$ model, a low-energy effective theory derived from the strong-coupling limit of the Holstein-Hubbard model, by large-scale variational Monte Carlo simulations. When the electron density is sufficiently low, the favored ground state is an s-wave PDW, consistent with results obtained from previous studies in this limit. Additionally, a PDW ground state with nematic d-wave pairing emerges in the intermediate range of electron densities and phonon frequencies. For these s-wave and d-wave PDWs arising in states with spontaneous breaking of time-reversal and inversion symmetries, PDW formation derives from valley-polarization and intra-pocket pairing.
\end{abstract}

\maketitle

{\bf Introduction:} A pair-density-wave (PDW) is an exotic superconducting (SC) state whose Cooper pairs have a nonzero center-of-mass momentum \cite{Agterberg2020}. While in a broad sense, the magnetic-field-induced Fulde-Ferrell-Larkin-Ovchinnikov (FFLO) states \cite{FF1964,LO1969} can be viewed as forms of PDW order, the term is usually used to refer to a state that is possible only in a strongly correlated electron fluid, typically without need of a magnetic field. It has been proposed that a  PDW phase arises in certain underdoped cuprate superconductors, for example, $1/8$ hole-doped LBCO and other members of the La-214 family \cite{Li2007,Tranquada2021,Kao2023,Berg2007,Berg2009,Tranquada2011,Tranquada2012} and in Bi$_2$Sr$_2$CaCu$_2$O$_{8+x}$ (Bi-2212) \cite{Lee2014,Hamidian2016, Yayu-NP2018, Davis-Science2019, Yayu-PRX2021}. Experimental evidence of PDW order has also been reported in other materials including certain Kagome metals\cite{HongjunGao-Science2021,Wang2022}, UTe$_2$ \cite{Madhavan2020,Madhavan2023,Liu2022}, and some Fe-based superconductors \cite{Fujita2023,JianWang2022}.

However, it has proven challenging to theoretically establish the existence of PDW ground states in any microscopic model, other than for certain 1D models \cite{Berg2010,Poilblanc2010,Fradkin2012,Mann2020,Zhang2023}, or in mean-field treatments of higher dimensional models in a strongly interacting regime where the validity of such solutions is largely uncertain \cite{Fradkin2009,Zhang2009,Kopp2011,Zhai2012,Moore2012,Fradkin2014,Fradkin2015,Chubukov2015,Yao2015,Hur2016,Granath2017,Granath2018,ZhouS2022,ZhouY2022,Fernandes2023,Santos2023,Raghu2023,WuYM023,Hirschfeld2023,Barlas2023,WuZ2023,WangYX2023,Castro2023,HanZY2023}. For instance, density-matrix renormalization group (DMRG) studies of various models on ladders have either found no evidence of significant PDW correlations, or at best of PDW correlations that fall off fast enough as a function of distance that the corresponding susceptibility is presumably non-divergent, even at $T\to 0$ \cite{Kim2019,Xu2019,Jiang2021,Peng2021,JiangHC2023,JiangYF2023,Devereaux2023,Yang2023,Sheng2023,WangHX2024}.

\begin{figure}[b]
\includegraphics[width=\linewidth]{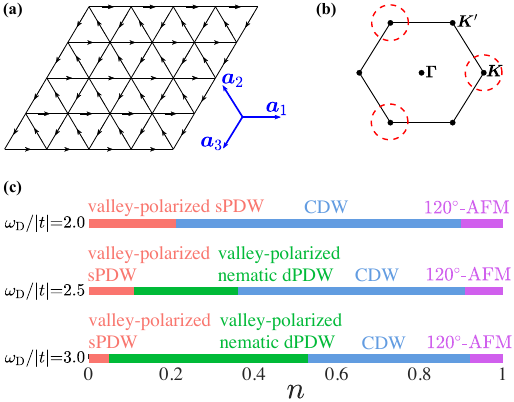}
\caption{(a) 2D triangular lattice where the arrows indicate the direction of the orbital currents in the valley-polarized ground states. (b) The first Brillouin zone is contained within the solid lines with $\pmb \Gamma$, $\pmb K$, $\pmb K'$ points marked. The red circles indicate the occupied valleys in the valley-polarized state. (c) Phase diagrams of the effective $t$-$J$-$V$ model in the limit of large system size with various electron densities, $n=N_e/N$. Here $\omega_{\text{D}}$ is phonon frequency in the Holstein-Hubbard model. The specific parameters used in the calculation are listed in Table~\ref{tab:parameter}. The PDW ordering vector is $\pmb Q=2\pmb K$. Abbreviations: sPDW = s-wave pair density wave; nematic dPDW = d-wave PDW plus broken lattice rotational symmetry.}
\label{fig:PhaseDiagram}
\end{figure}

Recently, three of us \cite{Han2020} established analytically that PDW order occurs in the 2D triangular lattice Holstein-Hubbard model in the strong-coupling limit for a narrow range of parameters. More generally, we showed that in the strong coupling limit, the low-energy physics is accurately represented by an effective model operating in a reduced subspace (with no double occupancy), that is similar to the $t$-$J$-$V$ model familiar as the strong-coupling limit of the plain Hubbard model. However, one new feature is that electron hoppings are suppressed by a Frank-Condon factor such that the effective model can be in an unfamiliar range of parameters (\emph{e.g.} $t \ll J, V$). In particular, the PDW was shown to emerge for dilute electrons, negative $t$ \cite{Han2020}, and an appropriate range of phonon frequencies. A DMRG study of the effective $t$-$J$-$V$ model on ladders with up to eight legs showed further evidence of a PDW  with valley-polarization for a larger range of parameters than could be treated analytically \cite{Huang2022}. However, further work is needed, even for the $t$-$J$-$V$ model, to establish the existence of a PDW groundstate in 2D and to extend the results beyond the dilute limit and beyond the narrowly constrained (and rather artificial) range of parameters that could be treated analytically.

We thus investigate an effective $t$-$J$-$V$ model on the 2D triangular lattice with up to $21 \times 21$ sites using a variational Monte Carlo (VMC) approach. Specifically, we consider a set of variational states with different patterns of broken symmetries with variable strengths of uniform (BCS) superconducting, PDW, charge-density-wave, nematic, and time-reversal symmetry breaking order parameters. We use VMC to identify the lowest variational energy among all these states. Despite known shortcomings of variational approaches to many-body systems \cite{AndersonBasicNotions}, such approaches often have proven successful in identifying qualitative aspects of many ground-state phase diagrams. Parameters for the effective $t$-$J$-$V$ model [Eq.~(\ref{t-J-V}) below] we have studied are derived from the original Holstein-Hubbard model (see Table~\ref{tab:parameter} below).

Our VMC study produces the quantum phase diagram shown in Fig. \ref{fig:PhaseDiagram}(c). Salient features are: (i) A s-wave PDW in the low electron density regions for $\omega_{\text{D}}/|t|=2$. Here, electrons are valley-polarized with spontaneous breaking of time-reversal and inversion symmetries, and the PDW arises from intra-pocket pairing with center-of-mass momentum $\pmb Q=2\pmb K$. (ii) For $\omega_{\text{D}}/|t|\gtrsim 2.5$, a distinct nematic d-wave PDW phase with broken lattice rotational symmetry arises at intermediate electron density. Both PDWs found in our study are commensurate which differs from the weakly incommensurate PDW found in the DMRG study \cite{Huang2022}; as discussed below, this disagreement may be real, but may be a finite size artifact of the present calculation. (iii) PDW order disappears when the electron density is increased beyond a critical value, beyond which charge density wave (CDW) and $120^{\circ}$ antiferromagnetic ($120^{\circ}$-AFM) order appear for all $\omega_{\text{D}}$ we studied.

{\bf Model and method:} In the present paper, we study the effective $t$-$J$-$V$ model on the triangular lattice:
\begin{align}\label{t-J-V}
H_{\text{eff}} = & - t_1 \sum_{\langle i,j\rangle,\sigma} \Big[c_{i\sigma}^{\dagger} c_{j\sigma} + {\rm H.c.} \Big ] \nonumber \\
& -  t_2 \sum_{\langle i, m, j\rangle,\sigma} \Big[  c_{i\sigma}^{\dagger} (1-2 n_m) c_{j\sigma} + {\rm H.c.} \Big ] \nonumber \\
& - (\tau+2t_2)\sum_{\langle i, m, j\rangle} \Big[ s^{\dagger}_{im} s_{mj} + {\rm H.c.} \Big ] \nonumber \\
& + J\sum_{\langle i,j\rangle }\left[\mathbf{S}_i\cdot\mathbf{S}_j - \frac{n_i n_j}{4} \right] + V \sum_{\langle i,j\rangle} n_{i} n_{j},
\end{align}
where projection onto states with no doubly occupied sites is implicit, $\langle i,j \rangle$ and $\langle i,m,j \rangle$ represent the sums over the nearest-neighbor sites and triplets of sites such that $m$ is a nearest-neighbor of two distinct sites $i$ and $j$, respectively. $c_{i\sigma}^\dag$ is a creation operator of spin $\sigma$ ($\uparrow$ or $\downarrow$) at site $i$, $\mathbf{S}_i= c_{i\alpha}^\dag\frac{\pmb\sigma_{\alpha\beta}}{2}c_{i\beta}$ is the spin operator on site $i$, $n_i=\sum_{\sigma}c^\dag_{i\sigma}c_{i\sigma}$ is the electron number operator, and $s_{ij} = \frac{1}{\sqrt{2}}(c_{i\uparrow} c_{j\downarrow}+c_{j\uparrow} c_{i\downarrow} )=s_{ji}$ is the annihilation operator of a singlet Cooper pair on bond $\langle i,j\rangle$.

The original Holstein-Hubbard model contains four independent parameters with units of energy, the bare nearest-neighbor hopping, $t$, the bare phonon frequency, $\omega_{\text{D}}$, the phonon induced attraction, $U_{\text{e-ph}}$, and the bare Hubbard $U_{\text{e-e}}$; the parameters that appear in the effective $t$-$J$-$V$ model are functions of these \cite{Han2020}. The values of these - and the bare parameters they come from - are then listed in Table \ref{tab:parameter}. In particular, we assume $U_{\text{e-e}} > U_{\text{e-ph}}\gg |t|$ which is what gives rise to the no double-occupancy condition. As can be seen from the table, for small $\omega_{\text{D}}$, the effective electron hoppings (\emph{e.g.} $t_1$, $t_2$, and $\tau$) are strongly suppressed while the exchange interaction $J$ and density-density interaction $V$ are not significantly affected. We have not considered very low phonon frequencies here because the effective electron hoppings will become exponentially smaller, so any PDW or SC states would be expected to have such low coherence scales that they are likely preempted by other types of order.

\begin{table}[t]
\vspace{-0.7em}
\caption{Parameters in the effective $t$-$J$-$V$ model [Eq. (\ref{t-J-V})] derived from the original Holstein-Hubbard model by fixing $t<0$, $U_{\text{e-e}}/|t|=22$, and $U_{\text{e-ph}}/|t|=18$ but varying the phonon frequency $\omega_{\text{D}}$ (see Ref. \cite{Han2020} for details).}
\centering
\begin{tabular}{c|c||cccc}
\hline
\hline
\hspace{0.1em} $\omega_{\text{D}}/|t|$ \quad & $J/|t|$ \quad & \quad $V/J$ \quad & \quad $t_1/J$ \quad & \quad $t_2/J$ \quad & \quad $\tau/J$  \\
\hline
\hspace{0.1em} 2.0  \qquad  & 0.20  \quad   & \quad 0.65  \quad   & \quad $-0.056$  \quad     & \quad 0.0081 \quad      & \quad 0.0096  \quad    \\
\hline
\hspace{0.1em} 2.5 \qquad & 0.20  \quad   & \quad 0.66  \quad   & \quad $-0.14$  \quad     & \quad 0.019  \quad     & \quad 0.024  \quad    \\
\hline
\hspace{0.1em} 3.0 \qquad   & 0.21  \quad   & \quad 0.67 \quad    & \quad $-0.24$  \quad     & \quad 0.033  \quad     & \quad 0.044  \quad    \\
\hline
\hline
\end{tabular}
\label{tab:parameter}
\end{table}

\begin{figure*}[t]
\includegraphics[width=\linewidth]{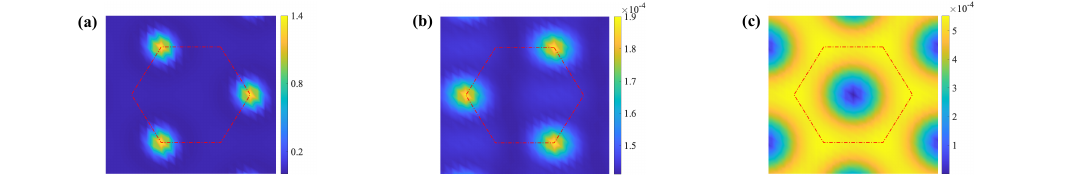}
\caption{Properties of the lowest-energy variational sPDW state for the model with $\omega_{\text{D}}/|t|=2$ and filling $n=1/12$. (a) The occupation number in momentum space $n(\pmb{k})=\sum_\sigma \langle c^\dag_{\pmb{k}\sigma}c_{\pmb{k}\sigma}\rangle$ as a function of $\pmb k$ in the first Brillouin zone (represented by the red dashed line) is plotted in color scale; it is clear that the electrons are valley polarized. (b) The static SC structure factor $|S^{\text{SC}}_{\pmb{a}_1,\pmb{a}_1}(\pmb{k})|$ plotted in color scale; 
(c) The static spin structure factor $S(\pmb{k})$ plotted in color scale. }
\label{fig:sPDW}
\end{figure*}

The VMC approach we adopt uses Gutzwiller projected trial wave functions, which is a powerful method to deal with strongly correlated systems \cite{Gros2007,Sorella2017,Imada2019}. Specifically, we use Gutzwiller-projected mean-field-type wave functions of the form $P_{\rm{N_e}}P_{\rm G}|\Psi_{\rm mf} \rangle$ where $P_{\rm G}$ projects out any states with doubly occupied sites, $P_{\rm {N_e}}$ is the projection operator onto states with a fixed number of electrons $N_e$, and  $| \Psi_{\rm mf} \rangle$ is constructed as the ground state of the following quadratic (mean-field-like) Hamiltonian \cite{Ogata2002,Imada2008},
\begin{align} \label{meanfield}
H_{\text{mf}} &= \sum_{i, l}\Big( T_{i,a_l} \mathcal{F}_i^\dag \mathcal{F}_{i+a_l} + \Delta_{i,a_l} \mathcal{F}_i^\dag \bar{\mathcal{F}}_{i+a_l} + {\rm H.c.} \Big) \nonumber \\
& ~~~ + \sum_i \Big( \Lambda_i  \mathcal{F}_i^\dag \mathcal{F}_i  + \mathbf{M}_i\cdot \mathcal{F}_i^\dag \frac{\pmb \sigma}{2} \mathcal{F}_i \Big ),
\end{align}
where $\mathcal{F}_i=(c_{i\uparrow}, c_{i\downarrow})^\mathsf{T}$, $\bar{\mathcal{F}}_i=(c_{i\downarrow}^\dag, -c_{i\uparrow}^\dag)^\mathsf{T}$, and $\pmb a_l$ with $l\in \{1,2,3\}$ are primitive vectors shown in Fig.~\ref{fig:PhaseDiagram}(a). Here $T_{i,a_l}$ are effective complex hopping amplitudes, $\Delta_{i,a_l}$ effective pair-fields, $\Lambda_i$ is a site energy, and $\mathbf{M}_i$ is an effective Zeeman field that induces various patterns of magnetic order. A pattern of phase order for the hopping amplitudes is assumed, as shown in Fig. \ref{fig:PhaseDiagram}(a), where the arrows indicate the direction of positive $\phi_{i,a_l}$ in $T_{i,a_l}=|T_{i,a_l}|e^{i\phi_{i,a_l}}$ with $\phi_{i,a_l}\in (-\pi,\pi$].

The various mean-field wave functions 
that we have analyzed correspond to different patterns and magnitudes of the variational parameters in Eq.~(\ref{meanfield}).  
A PDW state corresponds to a spatial varying pairing field,  
\begin{eqnarray}
\Delta_{i,a_l} =\left \{
 \begin{array}{ll}
   \tilde{\Delta}_{l} ( e^{i\pmb q \cdot \pmb r_i } + e^{-i\pmb q \cdot \pmb r_i } ),&\text{LO-type} \vspace{2mm}\\
   \tilde{\Delta}_{l}  e^{i\pmb q \cdot \pmb r_i },&\text{FF-type}
   \end{array}
\right.
\end{eqnarray}
where $\tilde{\Delta}_{l}$ represents pairing field on bonds along $\pmb{a}_l$ direction. Whenever $\pmb q=\pmb K$, which is the high-symmetry momentum point respecting $C_3$ rotational symmetry, we can set $\tilde{\Delta}_{l}=\Delta_s$, $\Delta_d \cos(2 \theta_l)$, or $\Delta_{d+id}e^{2i\theta_l}$ corresponding to s-wave PDW, or d-wave nematic PDW, d+id-wave PDW, respectively ($\theta_{l}$ being the azimuth angle of $\pmb{a}_l$). The s-wave PDW is invariant under $C_3$, while a d-wave PDW, in which $\theta_{l}$ can be $0$, $2\pi/3$, or $4\pi/3$, is necessarily nematic \cite{Ogata2004,Grover2010,MengCheng2010}. Note that $\pmb q$ is the center of mass momentum of the pairing in the mean-field wave function before Gutzwiller projection and it is a variational parameter. A translationally-invariant SC trial wave function corresponds to $\pmb q=0$. We also consider CDW order through modulated $\Lambda_i$, $120^{\circ}$-AFM order through three-sublattice local field $\mathbf{M}_i$, as well as states with co-existing multiple broken symmetries [for details see Supplemental Material (SM)].

In implementing the VMC calculations, the projections $P_{\rm {N_e}}$ and $P_{\rm G}$ must be enforced exactly.  The central quantity to be computed is the variational energy, $E(x) = \langle\Psi(x)|H_{\text{eff}}|\Psi(x)\rangle/\langle\Psi(x)|\Psi(x)\rangle$, where $|\Psi(x)\rangle = P_{\rm {N_e}}P_{\rm G}|\Psi_{\rm mf}(x)\rangle$, and $x$ denotes the variational parameters ($T_{i,a_l}$, $\Delta_{i,a_l}$, $\Lambda_i$, $\mathbf{M}_i$, $\pmb q$). The calculations have been performed on 2D triangular lattices with up to $21 \times 21$ sites and periodic boundary conditions along both directions (namely on a torus).

{\bf VMC results:} Here we study physics of the effective $t$-$J$-$V$ model whose parameters are listed in Table \ref{tab:parameter}. The main results are the VMC phase diagrams as a function of electron densities shown in Fig.~\ref{fig:PhaseDiagram}(c). For relatively small electron density, the ground state is a valley-polarized s-wave PDW state, and the region of sPDW decreases with increasing the phonon frequency. For a range of relatively large phonon frequencies, a d-wave PDW state arises at somewhat larger electron densities. The dPDW state is nematic in the sense that it breaks lattice rotational symmetry spontaneously. Both sPDW and dPDW ground states exhibit spontaneous valley polarization and hence break time-reversal and inversion symmetries. If the density of electrons is further increased, CDW and $120^{\circ}$-AFM states appear. Below we shall discuss in detail the properties of the PDW states.

\begin{table}[b]
\centering
\caption{Mean-field \textit{Ans\"atze} for different projected wave functions appearing in the phase diagrams, where $\theta_{l}$ is the azimuth angle of $\pmb{a}_l$. }
\begin{tabular}{l|c|c|c|c}
\hline
\hline
      & $T_{i,a_l}$  &  $\Delta_{i,a_l}$  &  $\Lambda_i$  &  $\mathbf{M}_i$  \\
\hline
\hline
sPDW  &  $ T $     & $\Delta_s e^{i\pmb q\cdot \pmb r_{i}}$    &  $\Lambda_0$       &      0\\
\hline
dPDW  & $T$        &  $\Delta_d \cos(2\theta_{l})e^{i\pmb q\cdot \pmb r_{i}} $  &  $\Lambda_0$       &     0       \\
\hline
CDW  & $ T $        &     0                            &  $\Lambda_0 + \Lambda_1 \cos(\pmb q \cdot \pmb r_i)$     &       0      \\
\hline
$120^{\circ}$    & $ T $          &       0            &   $\Lambda_0$     &    $120^{\circ}$-AFM        \\
\hline
\hline
\end{tabular}
\label{tab:Pattern}
\end{table}

\begin{figure*}[t]
\includegraphics[width=\linewidth]{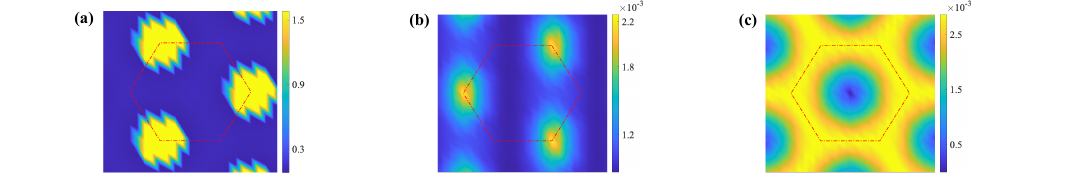}
\caption{Properties of the dPDW state with $\omega_{\text{D}}/|t|=3$, $n=5/12$. (a) The occupation number in momentum space $n(\pmb{k})$; (b) The static SC structure factor $|S^{\text{SC}}_{\pmb{a}_1,\pmb{a}_1}(\pmb{k})|$; (c) The static spin structure factor $S(\pmb{k})$. The red dotted box represents the first Brillouin zone.}
\label{fig:dPDW}
\end{figure*}

We find that for low electron density the Gutzwiller projected ground state with the lowest variational energy is always valley polarized. Valley polarization breaks inversion and time-reversal symmetries, but is consistent with translation symmetry. It also features an orbital loop current order of the sort shown in Fig.~\ref{fig:PhaseDiagram}(a) - which is a form of intra-unit cell orbital antiferromagnetism (or altermagnetism to use a new buzz word). Indeed, typically we find fully valley polarized states, so as shown in Fig.~\ref{fig:PhaseDiagram}(b) the Fermi surface encloses only one of two related high symmetry points, taken to be $\pmb K$ in the figure. In the presence of such order, if SC is to arise at all it is natural to expect PDW ordering since pairing two electrons in the same valley results in Cooper pairs with center-of-mass momentum $\approx 2\pmb K$. Indeed, as discussed below, although we treat the ordering vector $\pmb q$ as a variational parameter as shown in Table \ref{tab:Pattern}, so long as $\pmb q \approx \pmb Q \equiv 2 \pmb K$, after Gutzwiller projection we obtain a PDW {\em state} with ordering vector exactly equal to $\pmb Q$.

To see if the projected state exhibits valley polarization in the low electron density region, we calculate the electron occupation number in momentum space,
\begin{align}
n(\pmb{k})= \frac{1}{N}\sum_{i,j,\sigma}e^{i\pmb{k} \cdot (\pmb{r}_i-\pmb{r}_j)}\langle c_{i\sigma}^{\dagger}c_{j\sigma}\rangle=\sum_\sigma \langle c^\dag_{\pmb k\sigma}c_{\pmb{k}\sigma}\rangle
\end{align}
where $N$ is the number of sites of the lattice under study. As shown in Fig.~\ref{fig:sPDW}(a), essentially all of the electron occupation is concentrated at momenta around the $\pmb K$ valley, and the maximum value of $n(\pmb{k})$ is always larger than $1$. This feature is inherited directly from the pre-projected valley-polarized wave function \textit{ansatz}. Naturally, as the electron density increases, the effective Fermi surface of sPDW increases proportionately. (See the SM.)

To characterize the SC state, we calculate the static superconducting structure factor in the projected state,
\begin{align}
S^{\rm{SC}}_{\pmb a,\pmb a^\prime}(\pmb{k})= \frac{1}{N^2}\sum_{\pmb{r}_i,\pmb{r}_j}\mathrm{e}^{\mathrm{i}\pmb{k}\cdot (\pmb{r}_i-\pmb{r}_j})\langle \Delta^{\dagger}_{\pmb a}(\pmb{r}_i)\Delta_{\pmb a^\prime }(\pmb{r}_j)\rangle,
\end{align}
where $\Delta_{\pmb a}^\dag(\pmb{r}_i)=(c_{\pmb{r}_i,\uparrow}^\dag c_{\pmb{r}_i+\pmb{a},\downarrow}^\dag + c_{\pmb{r}_i+\pmb{a},\uparrow}^\dag c_{\pmb{r}_i,\downarrow}^\dag)$ creates a singlet pair on the nearest-neighboring bonds along the $\pmb{a}$ direction. We find that the SC structure factor peaks at momentum $\pmb Q=2\pmb K \equiv \pmb K'$, as shown in Fig.~\ref{fig:sPDW}(b), over the entire range of density we have explored and for any reasonable choice of  $\pmb q$ in the trial wave function before projection. We take this as indicating the existence of PDW order that arises from intra-pocket pairing with a commensurate center of mass momentum, as shown in Fig.~\ref{fig:PhaseDiagram}(b). As the electron density increases, the peak of the SC structure factor becomes higher (see the SM for details). Note that DMRG calculations of the model on cylinders with $L_x\gg L_y$ find an incommensurate s-wave PDW \cite{Huang2022}, which we attribute to the largeness of $L_x$ of the length of long cylinders studied by DMRG (see the SM for more details); namely we conjecture that the PDW obtained from VMC might be weakly incommensurate, consistent with the DMRG result, if the VMC study was extended to sufficiently large lattices.

To determine magnetic properties of the projected states, we compute the static spin structure factor $
S(\pmb{k})= \frac{1}{N^2}\sum_{ij}e^{i\pmb{k} \cdot (\pmb{r}_i-\pmb{r}_j)}\langle \pmb S_{\pmb{r}_i}\cdot \pmb S_{\pmb{r}_j}\rangle$. As shown in Fig. \ref{fig:sPDW}(c), for the sPDW state there appear no prominent peaks in the $S(\pmb{k})$. This is consistent with the observation that in this range of parameters, the variation energy is minimized when $\mathbf{M}_i=\mathbf{0}$. There is no long-range magnetic ordering in the sPDW state although the valley-polarization spontaneously breaks time-reversal symmetry.

As the PDW ordering vector is at high symmetry momentum point $\pmb K$, the internal structure of the PDW can be s-wave, d-wave, and even d+id. For $\omega_{\text{D}}/|t|=2$, we find that the energy of the optimized sPDW state is always lower than that of the dPDW state from low to intermediate electron density. However, for $\omega_{\text{D}}/|t|=2.5$ and $\omega_{\text{D}}/|t|=3$, we find that the dPDW state is lower for the large density end of the range we have explored, as shown in Fig.~\ref{fig:PhaseDiagram}(c). In common with what we found in the sPDW case, the variationally determined hopping parameters $T_{i,a_l}$ are such that time-reversal and inversion symmetries are spontaneously broken, and the electrons are valley polarized, as shown in Fig.~\ref{fig:dPDW}(a), and the superconducting state has pairing momentum $\pmb Q=2\pmb K \equiv \pmb K'$, as shown in Fig.~\ref{fig:dPDW}(b). The spin structure factor of the dPDW state is shown in Fig.~\ref{fig:dPDW}(c). We find that the region of dPDW increases with increasing phonon frequency, as shown in Fig.~\ref{fig:PhaseDiagram}(c). The dPDW state spontaneously breaks a discrete $C_3$ rotation symmetry of the triangular lattice, namely, it has nematic order. It is interesting to note that for the $t$-$J$-$V$ model studied here the nematic dPDW state has lower energy than the PDW with d+id pairing, which is also allowed by symmetry \cite{MengCheng2010,Grover2010}.

\begin{figure}[b]
\includegraphics[width=\linewidth]{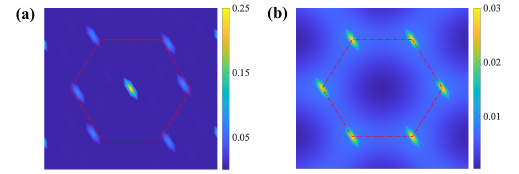}
\caption{(a) The density structure factor $C(\pmb k)$ of the CDW order with $\omega_{\text{D}}/|t|=2.5$, $n=0.5$; (b) The static spin structure factor $S(\pmb{k})$ of the $120^{\circ}$-AFM order with $\omega_{\text{D}}/|t|=2.5$, $n=0.94$. The red dotted box represents the first Brillouin zone.}
\label{fig:CCandSS}
\end{figure}

It is important to stress that the sign of the bare hopping $t$ in the Holstein-Hubbard model is quite crucial to induce PDW ground states, as anticipated in Ref. \cite{Han2020}. When $t>0$ (resulting in only a change in the sign of $t_1$ in the effective $t$-$J$-$V$ model), the ground state found in our VMC simulations at low electron density is a uniform d-wave SC rather than a PDW.

For completeness, we have cursorily explored the evolution of the ground-state to larger values of the electron density $n$, beyond the PDW ordering range.  Here, the typical ground state is a CDW without magnetic orders. The CDW is obtained from the projection of a mean-field state with spatial variations of on-site energy $\Lambda_i$, as shown in Table \ref{tab:Pattern}. We calculated CDW states with different momenta to obtain the state with the lowest possible energy. We find that the momentum $\pmb k^*$ of the optimized CDW state is typically incommensurate, with an amplitude that decreases when increasing $n$. The density structure factor $C(\pmb k)=\frac{1}{N^2}\sum_{\pmb{r}_i,\pmb{r}_j}e^{i\pmb{k} \cdot (\pmb{r}_i-\pmb{r}_j)}\langle n_{\pmb{r}_i}n_{\pmb{r}_j}\rangle$ is indeed peaked at the incommensurate $\pm \pmb k^*$ as well as at the $\pmb\Gamma$ point (which represents the uniform component of the charge density), as shown in Fig.~\ref{fig:CCandSS}(a). In other words, it is an incommensurate CDW.

At half filling ($n=1$), the effective model is equivalent to the spin-1/2 antiferromagnetic Heisenberg model on the triangular lattice whose ground state is believed to be the three-sublattice $120^{\circ}$ magnetically ordered state \cite{Huse1988,Huse1992,Sorella1999,Giamarchi2006,White2007}. Our VMC results for half-filling also show that the $120^{\circ}$-AFM order is the state with the lowest energy. For example, strong peaks at $\pmb K$ and $\pmb K'$ points emerge in the static spin structure factor $S(\pmb k)$ [Fig.~\ref{fig:CCandSS}(b)]. Furthermore, the $120^{\circ}$-AFM ordered state is stable when electron density is slightly away from half-filling although the single-particle excitation spectrum is gapless.

{\bf Discussions and conclusions:} The effects of relatively weak electron-phonon coupling are well understood - typically they lead to uniform SC or, in special circumstances, CDW order. Strong electron-phonon coupling, by contrast,  quenches electron mobility through polaron formation, when (as is typically the case) $\omega_{\text{D}}$ is small compared to the electronic bandwidth. This generally leads to insulating CDW order - possibly (in the presence of strong electron-electron repulsion) with coexisting magnetism. The present study deals with a relatively unexplored range of parameters, where the electron-phonon coupling is reasonably strong, but the phonon frequencies and electronic energies are comparable. While such a regime of parameters is unlikely to arise in most simple metals, it may be relevant in various artificial ``narrow-band'' systems such as 2D Moir{\'e} materials \cite{MacDonald2011,Cao2018a,Cao2018b} that have been the focus of considerable recent experimental interest, or in strongly correlated systems such as heavy-fermion systems where the electron effective mass can be highly renormalized. As Moir{\'e} systems or heavy-fermion systems have electron-phonon interactions, the strong electron-phonon coupling and the comparability between phonon frequencies and electron energies are, in principle, highly possible \cite{FWu2018,Lian2019,Choi2021,Young2019,Jorio2021}.

For PDW states of LO-type, secondary orders such as CDW and uniform charge-4e pairing can generically emerge, as can be understood from the perspective of  Ginzburg-Landau (GL) theory \cite{Berg2009,Tsunetsugu2008,Tranquada2015}. Our VMC calculations suggest that the ground state at low electron densities in an appropriate range of phonon frequencies is a PDW state of FF-type, namely a valley-polarized PDW state (s-wave or d-wave) whose Cooper pairs condense at a single momentum $\pmb Q=2\pmb K$. Consequently, there is no induced CDW or uniform charge-4e SC order. However, there are higher harmonic orders consisting of charge-4e PDW order with $2\pmb Q$ and charge-6e uniform SC order with $3\pmb Q\equiv \pmb 0$ since $3\pmb K$ is a reciprocal lattice vector.

In summary, our VMC calculations imply that (s-wave or d-wave) PDW ground state phases occur in the Holstein-Hubbard model on the triangular lattice for moderately strong doping, relatively high phonon frequencies, in a range of (low) electron densities that depends on the phonon frequency, as shown in Fig.~\ref{fig:PhaseDiagram}(c). The obtained PDW state is valley-polarized with staggered loop currents, spontaneously breaking both time-reversal and inversion symmetries, which can be employed to realize superconducting diode effect \cite{Dai2007,Ono2020,Hu2022,Ali2022}. One interesting open question concerns the nature of the quantum phase transition between sPDW and dPDW. Moreover, how these different PDWs respond to external magnetic fields is another interesting issue for investigation \cite{Han2022}. We believe that our work provides strong support for possible realization of various PDW states in 2D quantum systems with both strong electron-electron and electron-phonon interactions.

{\it Acknowledgement:} This work was supported in part by the NSFC under Grant No.~12347107. JW acknowledges support from the NSFC under Grant No.~12404170 and the start-up grant at HZNU. ZH and SAK were funded in part by the Department of Energy, Office of Basic Energy Sciences, under Contract No.~DE-AC02-76SF00515 at Stanford. HY was supported in part by the New Cornerstone Science Foundation through the Xplorer Prize.

% \newpage

\begin{widetext}

\renewcommand{\theequation}{S\arabic{equation}}
\setcounter{equation}{0}
\renewcommand{\thefigure}{S\arabic{figure}}
\setcounter{figure}{0}
\renewcommand{\thetable}{S\arabic{table}}
\setcounter{table}{0}

\section{Supplemental Material}

\subsection{A. Different \textit{Ans\"atze} and competing states}\label{SM:energy}
In the following, we will show more Gutzwiller-projected mean-field states with different pairing patterns in the VMC framework. As shown in the main text, the general mean-field-like Hamiltonian (with spin-singlet pairing terms) can be written as,
\begin{align}\label{SM:meanfield}
H_{\text{mf}} = &\sum_{i, l}\Big( T_{i,a_l} \mathcal{F}_i^\dag \mathcal{F}_{i+a_l} + \Delta_{i,a_l} \mathcal{F}_i^\dag \bar{\mathcal{F}}_{i+a_l} + {\rm H.c.} \Big)
+ \sum_i \Big( \Lambda_i  \mathcal{F}_i^\dag \mathcal{F}_i  + \mathbf{M}_i\cdot \mathcal{F}_i^\dag \frac{\pmb \sigma}{2} \mathcal{F}_i \Big ), \\
&{\rm where} \qquad \Delta_{i,a_l} =\left \{
 \begin{array}{ll}
   \tilde{\Delta}_{l} ( e^{i\pmb q \cdot \pmb r_i } + e^{-i\pmb q \cdot \pmb r_i } ), &\text{LO-type} \vspace{2mm}\\
   \tilde{\Delta}_{l}  e^{i\pmb q \cdot \pmb r_i },&\text{FF-type}
   \end{array}
\right.\nonumber
\end{align}
where $|\Psi_{\rm mf}(x)\rangle$ is the mean-field ground state, and $x$ denotes the variational parameters ($T_{i,a_l}$, $\Delta_{i,a_l}$, $\Lambda_i$, $\mathbf{M}_i$, $\pmb q$). Using the above trial wave functions and the Gutzwiller projection enforcing the no-double-occupancy constraint (i.e. $P_{\rm G}=\prod_i(1-n_{i,\uparrow}n_{i,\downarrow})$), we could calculate the expectation values of the original effective $t$-$J$-$V$ model employing the standard VMC approach. Specifically, the projected trial wave function becomes $|\Psi(x)\rangle = P_{\rm {N_e}}P_{\rm G}|\Psi_{\rm mf}(x)\rangle$ with fixed number of electrons $N_e$ through $P_{\rm {N_e}}$.

The PDW state is constructed by spatial variations of the pairing parameter, $\Delta_{i,a_l}$ with $l \in \{1,2,3 \}$. Note that $\pmb q$ is the center-of-mass momentum of the pairing in the mean-field wave function.
Indeed, from a symmetry point of view, (i) if the PDW ordering vector is at the high symmetry point (i.e. $\pmb \Gamma$, $\pmb K$, or $\pmb K'$), the classification of s- and d-wave is meaningful, and then the s-wave PDW is invariant under $C_6$ rotation while the d-wave PDW is necessarily nematic due to local structure; (ii) the rotation symmetry is necessarily broken even in an s-wave pairing if the PDW ordering vector $\pmb q$ is generic not at high symmetry point. Thus, there is no constraint on $\Delta_{i,a_l}$ for the generic momentum $\pmb q$, meaning that there is no relationship between the strength of the pairing among three distinct bond directions, namely a generic PDW. In other words, for the generic momentum $\pmb q$, the sPDW and dPDW are just two special choices of the former factors. For $\pmb q=\pmb K$ or $\pmb K'$, it has a $C_3$ rotation symmetry such that one can sharply define sPDW and dPDW. We have also constructed uniform SC states (with setting $\pmb q=0$), CDW (through modulated $\Lambda_i$), $120^{\circ}$-AFM order (via three-sublattice local field $\mathbf{M}_i$), and other mixed states with coexisting multiple broken symmetries (such as uniform SC+PDW states, uniform SC+CDW states, PDW+CDW states, uniform SC+$120^{\circ}$-AFM states, CDW+$120^{\circ}$-AFM states, and the like). In addition, we have constructed spin-triplet SC trial wave functions, but their energies are much higher than those of spin-singlet SC states, and thus we do not consider them here.

The details of the VMC method are briefly summarized below. The optimization of the variational parameters $x$ is achieved by minimizing the energy $E(x) = \langle\Psi|H_{\rm eff} |\Psi(x)\rangle /\langle\Psi(x)|\Psi(x)\rangle$, using the Nelder-Mead simplex search method \cite{Lagarias1998} and the stochastic reconﬁguration method \cite{Sorella1998,Capriotti2000}. The simplex search method is a direct search method that does not use gradients (indicating a derivative-free method), and it is efficient and stable for a few variational parameters. In order to speed up the convergence of Monte Carlo sampling, we adopt the Metropolis importance sampling approach for updating the configurations. In the simulation, we will often do the computation of the ratio of the determinants or pfaffians of two matrices. For the sake of efficiency, the algorithm that computes this quantity by storing the inverse of the old matrix is always used \cite{Ceperley1977}. Furthermore, we have performed several runs with randomly chosen initial configurations to independently check the error bars. In addition, to search for the possible global minimum, we have chosen many different initial variational parameters in each variational process.

\begin{table*}[t]
\caption{The optimized energy of several candidate states of the effective $t$-$J$-$V$ model obtained on a system of size $12 \times 12$ with various electric densities. Here we don't show $120^{\circ}$-AFM order (due to almost zero $\mathbf{M}_i$ after Gutzwiller projection) and possible mixed states (because of higher variational energies). Note that the optimized energy means the total energy of the original model instead of the energy per site and we set $|t| = 1$ as an energy unit.}
\centering
\begin{tabular}{l|ccc|ccc|ccc}
\hline
\hline
                        &  \multicolumn{3}{c|}{ $\omega_{\text{D}}/|t|=2.0$} & \multicolumn{3}{c|}{ $\omega_{\text{D}}/|t|=2.5$} &  \multicolumn{3}{c}{ $\omega_{\text{D}}/|t|=3.0$}  \\
                        & $n=1/6$  & \quad $n=1/4$  \quad &   $n=1/3$  \quad   &  $n=1/12$  & \quad $n=1/6$ \quad &  $n=5/12$  \quad   &  $n=1/4$   & \quad $n=1/3$ \quad &  $n=5/12$  \\
\hline
\multirow{2}{*}{Uniform s-wave SC}        & $-0.7287$   & $-0.7599$  &  $-0.2898$     & $-1.0018$  & $-1.6425$  & $-0.5895$       & $-3.4833$  & $-3.4173$  &  $-2.3484$ \\
                        & $\pm 0.0008$  &   $\pm 0.0010$   &  $\pm 0.0014$  &  $\pm 0.0003$  & $\pm 0.0007$  & $\pm 0.0020$  & $\pm 0.0010$ & $\pm 0.0016$ &  $ \pm 0.0020$ \\
\hline
\multirow{2}{*}{Valley-polarized sPDW}   & $-0.7878$   & $-0.7837$  &  $-0.3048$     & $-1.0116$  & $-1.6408$  & $-0.5904$  & $-3.4818$   & $-3.4182$  & $-2.3573$ \\
                        & $\pm 0.0010$  &   $\pm 0.0012$   &  $\pm 0.0014$  &  $\pm 0.0003$  & $\pm 0.0007$  & $\pm 0.0023$  & $\pm 0.0013$ & $\pm 0.0018$ & $\pm 0.0019$ \\
\hline
\multirow{2}{*}{Valley-unpolarized sPDW} & $-0.7738$   & $-0.7788$  &  $-0.2706$     & $-1.0103$  & $-1.6322$  &  $-0.5613$      & $-3.4816$  & $-3.3980$  & $-2.3557$ \\
                        & $\pm 0.0011$  &   $\pm 0.0012$   &  $\pm 0.0015$  &  $\pm 0.0003$   & $\pm 0.0007$  & $\pm 0.0021$  & $\pm 0.0011$ & $\pm 0.0016$ & $\pm 0.0021$ \\
\hline
\multirow{2}{*}{Uniform d-wave SC}       & $-0.7292$   & $-0.7588$  &  $-0.3062$     & $-1.0009$  & $-1.6436$  &  $-0.5941$      & $-3.4845$  & $-3.4103$  & $-2.3633$  \\
                        &  $\pm 0.0007$  &  $\pm 0.0011$   &  $\pm 0.0015$  &  $\pm 0.0003$  & $\pm 0.0007$   & $\pm 0.0019$     & $\pm 0.0012$ & $\pm 0.0016$ & $\pm 0.0021$ \\
\hline
\multirow{2}{*}{Valley-polarized dPDW}   & $-0.7292$   & $-0.7603$  &  $-0.3085$     & $-1.0008$  & $-1.6441$  &  $-0.5881$  & $-3.4867$  & $-3.4209$   & $-2.3675$ \\
                        &  $\pm 0.0007$  & $\pm 0.0010$    &  $\pm 0.0016$  &  $\pm 0.0003$   & $\pm 0.0007$  & $\pm 0.0019$     & $\pm 0.0012$ & $\pm 0.0017$ & $\pm 0.0021$ \\
\hline
\multirow{2}{*}{Valley-unpolarized dPDW} & $-0.7146$   & $-0.7394$  &  $-0.2882$     & $-0.9476$  & $-1.5717$  &  $-0.4153$      & $-3.4537$  & $-2.9171$  & $-2.1150$ \\
                        &  $\pm 0.0016$  &  $\pm 0.0011$     & $\pm 0.0017$ &  $\pm 0.0006$   & $\pm 0.0010$  & $\pm 0.0016$     &  $\pm 0.0015$ & $\pm 0.0107$ & $\pm 0.0030$ \\
\hline
\multirow{2}{*}{Uniform d+id-wave SC}    & $-0.7275$   & $-0.7561$  &  $-0.3057$     & $-1.0012$  &  $-1.6424$ &  $-0.5875$      & $-3.4851$  & $-3.4114$  & $-2.3579$ \\
                        &  $\pm 0.0008$  &   $\pm 0.0011$    &  $\pm 0.0016$ & $\pm 0.0003$   &  $\pm 0.0007$  & $\pm 0.0020$     & $\pm 0.0012$ & $\pm 0.0015$ & $\pm 0.0021$ \\
\hline
\multirow{2}{*}{Valley-polarized d+idPDW} & $-0.7255$   & $-0.7577$  &  $-0.3038$     & $-1.0007$  & $-1.6415$  &  $-0.5789$      & $-3.4811$  & $-3.4102$  & $-2.3539$ \\
                        &  $\pm 0.0008$  &   $\pm 0.0010$    & $\pm 0.0015$   & $\pm 0.0003$   &  $\pm 0.0007$  & $\pm 0.0019$     & $\pm 0.0012$ & $\pm 0.0016$ & $\pm 0.0023$ \\
\hline
\multirow{2}{*}{Valley-unpolarized d+idPDW} & $-0.7098$   & $-0.7042$  &  $-0.3037$     & $-0.9816$  &  $-1.5998$ &  $-0.3983$      & $-3.4583$  &  $-2.8440$ & $-2.2573$ \\
                        &  $\pm 0.0018$   &  $\pm 0.0013$    &  $\pm 0.0029$   & $\pm 0.0004$  &  $\pm 0.0008$  & $\pm 0.0025$     & $\pm 0.0012$ & $\pm 0.0020$ & $\pm 0.0163$ \\
\hline
\multirow{2}{*}{CDW}                    & $-0.6596$   & $-0.8876$  &  $-0.5466$     & $-0.9175$   &  $-1.4568$   & $-0.7589$   & $-2.8914$     & $-3.1204$    &  $-1.8722$ \\
                        &  $\pm 0.0008$  &  $\pm 0.0026$   &  $\pm 0.0013$  & $\pm 0.0004$ & $\pm 0.0008$ &  $\pm 0.0018$   & $\pm 0.0014$  & $\pm 0.0016$  & $\pm 0.0026$ \\
\hline
\hline
\end{tabular}\label{SM:energyall}
\end{table*}

Based on the VMC calculations, a direct comparison of the optimized energies of several candidate states of some typical parameter sets is shown in Table~\ref{SM:energyall}. As can be seen in the table, there is a phase transition (from valley-polarized sPDW to CDW) between electron densities $n=1/6$ and $n=1/4$ for $\omega_{\text{D}}/|t|=2$. Thus, the favored ground state is the s-wave PDW when the electron density is sufficiently low, consistent with the results of the previous perturbation approach \cite{Han2020}. And for $\omega_{\text{D}}/|t|=2.5$, the energy of the valley-polarized sPDW state is slightly lower than that of the d-wave PDW for $n=1/6$. Therefore, there is a direct phase transition from the sPDW to the dPDW below the electron density $n=1/6$. As the phonon frequency increases, there is a much larger region for the dPDW phase for $\omega_{\text{D}}/|t|=3$. Here, we also compare the energies of valley-polarized PDW states with different momentums $\pmb qs$ and find that their energies are no better (not shown here). From our VMC simulations, the optimized hopping parameters $T_{i,a_l}$ are complex numbers (which usually break time-reversal and inversion symmetries), and the symmetry between momenta $\pmb k$ and $-\pmb k$ points in the dispersion is indeed broken. In the case of a small electron density, the electrons will first occupy the momentum points where the energies are lower (i.e. valley polarization near the $\pmb K$ point in Fig.~1(b) of the main text), and the Gutzwiller projection can spontaneously select out the pairing momentum $\pmb Q = 2\pmb K \equiv \pmb K'$. Consequently, the mechanism of PDWs in the present model is first valley polarization with staggered loop currents and then intra-pocket pairing.

Actually, in the VMC framework, we have considered generic PDW states (namely $\tilde\Delta_l$ are considered as three independent variational parameters), and found that their energies are no better than those of the corresponding s-wave or d-wave PDW in the larger lattice size. Additionally, the effect of the small $t_2$ and $\tau$ terms in the effective $t$-$J$-$V$ model on the 2D triangular lattice is almost negligible, at most slightly changing the phase boundary. In summary, the hopping $t_1$, the exchange interaction $J$, and the density-density interaction $V$ determine the key physics of the system under study.

At half-filling, the effective model is equivalent to the spin-1/2 AFM Heisenberg model on the triangular lattice and the three-sublattice $120^{\circ}$ magnetically ordered state is the state with the lowest energy. In the present work, the $120^{\circ}$-AFM ordered state is shown to be stable when the electron density is only slightly away from half-filling. Phase separation (PS) away from half-filling is possible when slightly doping the $120^{\circ}$-AFM ordered state \cite{Ohgoe2017,Becca2017,Johnston2022}. As can be seen in Table I of the main text, the exchange interaction $J$ or density-density interaction $V$ is much larger than each effective electron hopping (e.g. $t_1$, $t_2$, and $\tau$), and then it is possible for the loss of the magnetic contribution of the total energy to be larger than the gain due to the kinetic energy. However, since the main interest in the present manuscript is in regions with low electron density where the PDW phase may emerge, 
we did not include the PS ansatz in our VMC calculations for the model slightly away from half-filling, which is left for future study.
Moreover, more accurate energies can be found by adding further variational parameters in the variational wave function, such as allowing for the Jastrow factor (adding further electron-electron correlations). However, in the present study we did not implement a Jastrow factor in the variational wave function, partly because the Jastrow factor will substantially increase the computational time for the VMC calculation with very large system sizes. In the future, we will try to explore the quantum phase diagrams of the triangular lattice $t$-$J$-$V$ model using wave functions with more variational parameters such as the Jastrow factors.

Taken together, our VMC simulations clearly show that the valley-polarized (either s-wave or d-wave) PDW state can be favored as the ground state of the triangular lattice $t$-$J$-$V$ model [see Fig.~1(c) in the main text]. It is possible that the sPDW state disappears with the further increase in the phonon frequency. However, further studies are required to determine the upper critical phonon frequency at which the valley-polarized sPDW disappears.

\begin{figure*}[t]
\includegraphics[width=\linewidth]{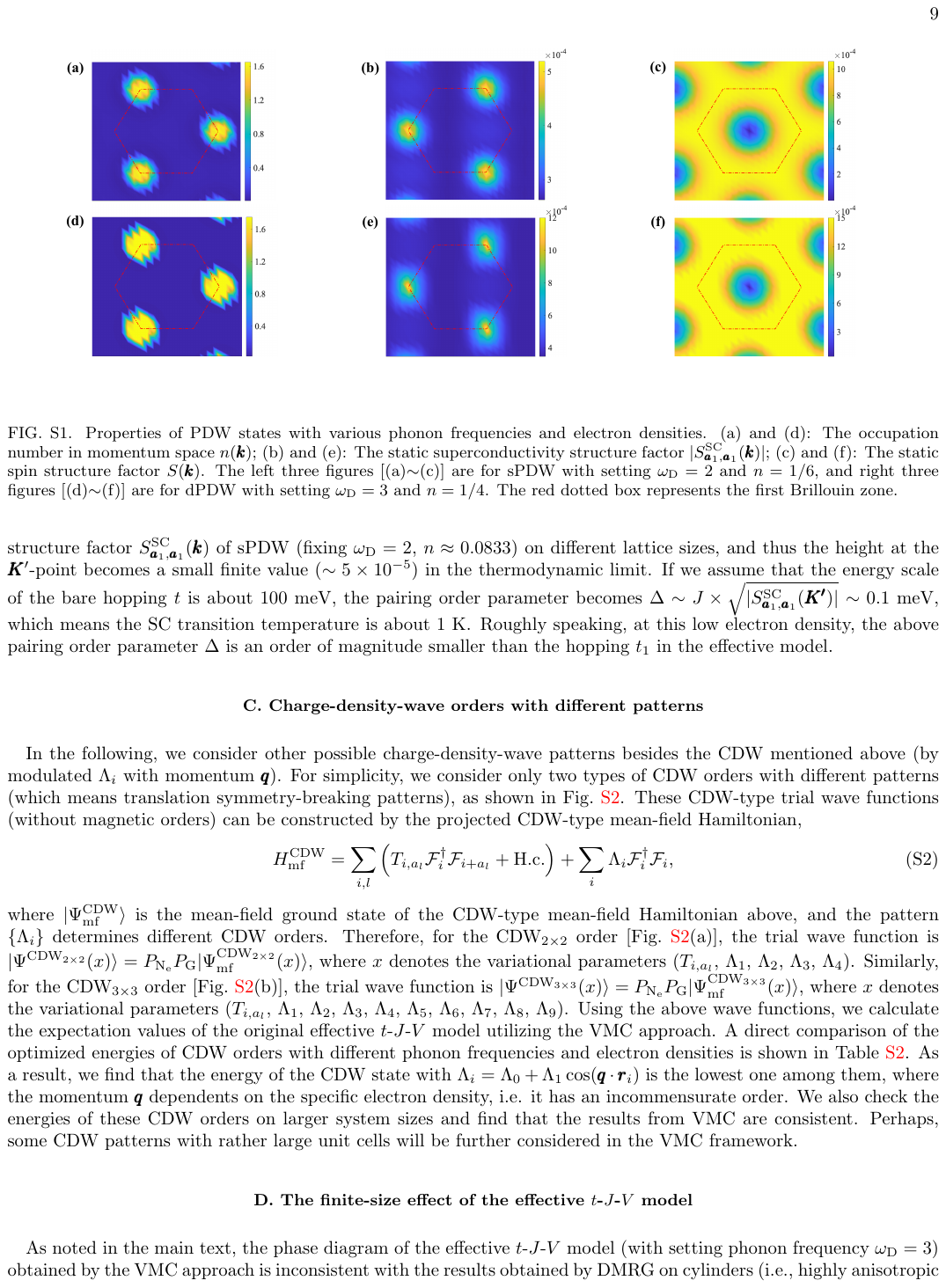}
\caption{Properties of PDW states with various phonon frequencies and electron densities. (a) and (d): The occupation number in momentum space $n(\pmb{k})$; (b) and (e): The static SC structure factor $|S^{\text{SC}}_{\pmb a_1, \pmb a_1}(\pmb{k})|$; (c) and (f): The static spin structure factor $S(\pmb{k})$. The left three figures [(a)$\sim$(c)] are for sPDW with setting $\omega_{\text{D}}/|t|=2$ and $n=1/6$, and right three figures [(d)$\sim$(f)] are for dPDW with setting $\omega_{\text{D}}/|t|=3$ and $n=1/4$. The red dotted box represents the first Brillouin zone.}
\label{SM:PDWs}
\end{figure*}

\subsection{B. Properties of PDWs with various electron densities}\label{SM:PDW}
As shown in Figs.~2, 3, and \ref{SM:PDWs}, the properties of PDW states are stable for electron densities regardless of whether the ground state is a valley-polarized sPDW or dPDW state. Specifically, for both PDW states, the static spin structure factor remains qualitatively unchanged while the effective Fermi surface (illustrated by the occupation number in momentum space $n(\pmb k)$) becomes larger and larger as we increase electron densities within the PDW phase. Furthermore, the peak of the static SC structure factor in the PDW phase becomes higher as the electron density increases. 
Consequently, the properties of the sPDW (or dPDW) are consistent in the corresponding region of the phase diagrams. In particular, the valley polarization (indicating an orbital loop current order) breaks time-reversal and inversion symmetries independently for the SC order. Conversely, the valley-polarized sPDW does not break any other symmetries, while the valley-polarized dPDW breaks the remaining $C_3$ rotation symmetry, namely, exhibiting a nematic order.

Below we briefly discuss why the PDW state is more stable than the uniform d-wave SC state. In the sense, the mechanism of the PDW state can be understood from the perspective of perturbation theory in the dilute limit of electrons and in the adiabatic limit of phonons (for more details see Ref.~\cite{Han2020}). 
Furthermore, an important question is whether the favored ground state is still a PDW beyond the adiabatic limit ($\omega_{\rm D} \rightarrow 0$) and beyond the dilute electron limit ($n \rightarrow 0$). From the large-scale VMC simulations, we find that for low electron density, the Gutzwiller projected ground state with the lowest variational energy is always valley polarized. It is important to note that valley polarization breaks inversion and time-reversal symmetries, yet it remains consistent with translational symmetry. In most cases, fully valley-polarized states are observed, with the Fermi surface enclosing only one of two high-symmetric points $\pm\pmb K$, as illustrated in Fig.~1(b) of the main text. In the presence of such valley polarization, if SC is to arise it is natural to expect PDW ordering since pairing two electrons in the same valley results in Cooper pairs with center-of-mass momentum $\approx 2\pmb K$. Indeed, the PDW state (deriving from valley polarization and intra-pocket pairing) is more stable than the uniform d-wave SC state.

\begin{figure*}[b]
\includegraphics[width=\linewidth]{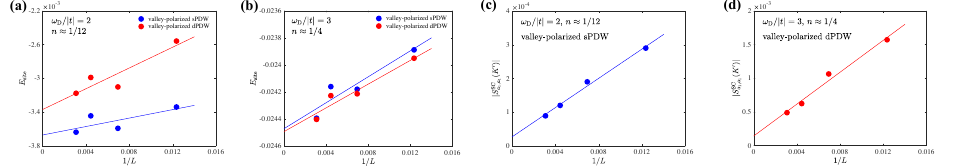}
\caption{The optimized ground-state energy for two candidate states of the effective $t$-$J$-$V$ model on the torus obtained with the parameter $\omega_{\rm D}/|t|=2$, $n\approx1/12$ (a) and $\omega_{\rm D}/|t|=3$, $n\approx1/4$ (b). The static SC structure factor $|S^{\text{SC}}_{a_1, a_1}(K')|$ for the corresponding valley-polarized sPDW (c) and dPDW (d). Note that the optimized energy means the energy per site, and the energy error bar is smaller than the symbol size. The linear fitting is used here.}\label{SMFiniteSizeScaling}
\end{figure*}

The following presents a finite-size analysis of energies and the static superconducting (SC) structure factor for a couple of fixed electron densities, as shown in Fig.~\ref{SMFiniteSizeScaling}.
For $\omega_{\text{D}}/|t|=2$, the energy of the valley-polarized sPDW state is lower than that of the dPDW at $n\approx 1/12$. In contrast, for $\omega_{\text{D}}/|t|=3$, the energy of the valley-polarized dPDW state is slightly lower than that of the valley-polarised sPDW at $n\approx 1/4$. Note that the energy oscillates slightly with system size, due to the fact that the exact electron density $n$ cannot always be realized on finite lattices-instead, the closest achievable value is used.
In addition, to estimate the SC transition temperature at low electron density, we try to extrapolate the static SC structure factor $|S^{\text{SC}}_{\pmb a_1, \pmb a_1}(\pmb K')|$ of the valley-polarized sPDW or dWave to the thermodynamic limit. The pairing order parameter is $\Delta \sim J\times \sqrt{|S^{\text{SC}}_{\pmb a_1, \pmb a_1}(\pmb{K'})|}$ for simplicity, which is about $0.01J$ for the d-wave PDW.

\begin{figure*}[t]
\includegraphics[width=\linewidth]{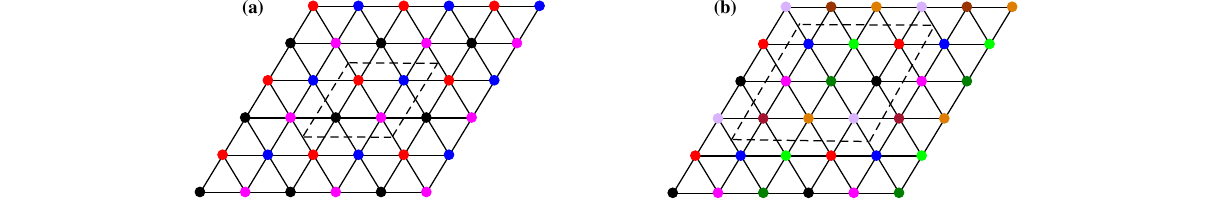}
\caption{The patterns of two types of CDW orders. (a) CDW$_{2\times 2}$ order with a $2\times 2$ pattern; (b) CDW$_{3\times 3}$ order with a $3\times 3$ pattern. Note that different colors indicate different site-energy parameters $\Lambda_i$ in Eq. (\ref{SM:CDW}). The black dotted box represents the unit cell of the corresponding CDW order. }
\label{SM:CDWs}
\end{figure*}

\subsection{C. Charge-density-wave orders with different patterns}
In the following, we consider other possible charge-density-wave (CDW) patterns besides the CDW mentioned above (by modulated $\Lambda_i$ with momentum $\pmb q$, as shown in Table II of the main text). For the sake of simplicity, we consider only two types of CDW orders with different patterns (i.e. translation symmetry breaking patterns), as shown in Fig.~\ref{SM:CDWs}. These CDW-type trial wave functions (without magnetic orders) can be constructed from the projected CDW-type mean-field Hamiltonian,
\begin{align}\label{SM:CDW}
H_{\text{mf}}^{\text{CDW}} &= \sum_{ i,l}\Big( T_{i,a_l} \mathcal{F}_i^\dag \mathcal{F}_{i+a_l} + {\rm H.c.} \Big)
  + \sum_i  \Lambda_i  \mathcal{F}_i^\dag \mathcal{F}_i,
\end{align}
where $|\Psi_{\rm mf}^{\rm{CDW}}\rangle$ is the mean-field ground state of the CDW-type mean-field Hamiltonian above, and the pattern $\{\Lambda_i \}$ determines different CDW orders. Therefore, for the CDW$_{2\times 2}$ order [Fig.~\ref{SM:CDWs}(a)], the trial wave function is $|\Psi^{\rm{CDW_{2\times 2}}}(x)\rangle = P_{\rm {N_e}}P_{\rm G}|\Psi_{\rm mf}^{\rm{CDW_{2\times 2}}}(x)\rangle$, where $x$ denotes the variational parameters ($T_{i,a_l}$, $\Lambda_1$, $\Lambda_2$, $\Lambda_3$, $\Lambda_4$). Similarly, for the CDW$_{3\times 3}$ order [Fig.~\ref{SM:CDWs}(b)], the trial wave function is $|\Psi^{\rm{CDW_{3\times 3}}}(x)\rangle = P_{\rm {N_e}}P_{\rm G}|\Psi_{\rm mf}^{\rm{CDW_{3\times 3}}}(x)\rangle$, where $x$ denotes the variational parameters ($T_{i,a_l}$, $\Lambda_1$, $\Lambda_2$, $\Lambda_3$, $\Lambda_4$, $\Lambda_5$, $\Lambda_6$, $\Lambda_7$, $\Lambda_8$, $\Lambda_9$). Using the above wave functions, we calculate the expectation values of the original effective $t$-$J$-$V$ model utilizing the VMC approach. A direct comparison of the optimized energies of CDW orders with different phonon frequencies and electron densities is shown in Table~\ref{SM:tabCDW}. As a result, we find that the energy of the CDW state with $\Lambda_i=\Lambda_0 + \Lambda_1 \cos(\pmb q \cdot \pmb r_i)$ is the lowest one among them, where the momentum $\pmb q$ is dependent on the specific electron density, i.e. it has an incommensurate order. We also checked the energies of these CDW orders at larger system sizes and found that the results obtained from VMC simulations are consistent. Perhaps some CDW patterns with substantial unit cells are worthy of further consideration within the VMC framework.

\begin{table}[t]
\caption{The optimized energy of CDWs obtained on a system of size $12 \times 12$ with various phonon frequencies and electron densities. The patterns of CDW$_{2\times 2}$ and CDW$_{3\times 3}$ are shown in Fig.~\ref{SM:CDWs} above. Note that the optimized energy means the total energy of the original model instead of the energy per site and we set $|t| = 1$ as an energy unit.}
\centering
\begin{tabular}{l|c|c|c}
\hline
\hline
      & \quad $\omega_{\text{D}}/|t|=2.0$, $n=1/3$     \quad  & \quad $\omega_{\text{D}}/|t|=2.5$, $n=5/12$  \quad  & \quad $\omega_{\text{D}}/|t|=3.0$, $n=5/9$   \\
\hline
CDW                & $-0.5466 \pm 0.0013$   & $-0.7589 \pm 0.0018$    & $0.3069 \pm 0.0029$   \\
\hline
CDW$_{2\times 2}$  & $-0.3156 \pm 0.0014$   & $-0.5962 \pm 0.0020$    & $0.6514 \pm 0.0024$    \\
\hline
CDW$_{3\times 3}$  & $-0.3140 \pm 0.0015$   & $-0.5974 \pm 0.0021$    & $0.6539 \pm 0.0025$     \\
\hline
\hline
\end{tabular}\label{SM:tabCDW}
\end{table}

\begin{figure*}[t]
\includegraphics[width=\linewidth]{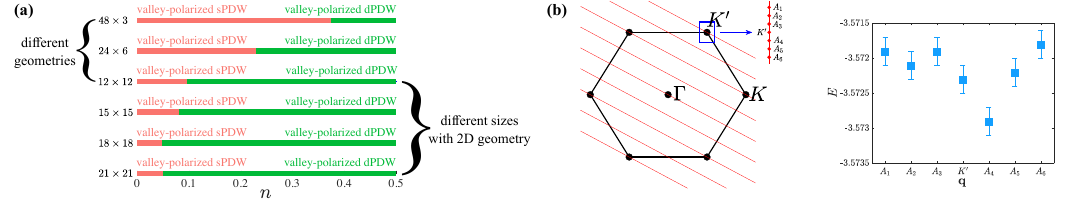}
\caption{(a) The VMC phase diagrams of the effective $t$-$J$-$V$ model (at fixed $\omega_{\rm D}/|t|=3$) with various electron densities $n$ in various lattice sizes. The lattice geometry is chosen to be an $L_x\times L_y$ torus with $48 \times 3$, $24 \times 6$, $12 \times 12$, $15 \times 15$, $18 \times 18$, and $21 \times 21$. Note that different geometries and sizes are considered to address the finite-size effect in the effective lattice model. (b) The first Brillouin zone with the red bands indicating allowed values of momenta for the torus with $72\times 6$. The optimized energies of the valley-polarized sPDW with a few nearby momentum $\pmb qs$ around high-symmetric $\pmb K'$-point, obtained with $\omega_{\text{D}}/|t|=3$ and $n\approx 0.05$.}
\label{SM:finite}
\end{figure*}

\subsection{D. The finite-size effect of the effective $t$-$J$-$V$ model}\label{SM:scaling}
As noted in the main text, the phase diagram of the effective $t$-$J$-$V$ model (with setting $\omega_{\text{D}}/|t|=3$) obtained by VMC is inconsistent with the results obtained by DMRG on cylinders (i.e., highly anisotropic ladder geometries composed of a relatively small number of legs)  \cite{Huang2022}. The DMRG results show that the ground state is only a valley-polarized sPDW (resulting from intra-pocket pairing with a weakly incommensurate center of mass momentum) in a large region of low electron density, while the VMC calculations show a valley-polarized sPDW for a small region with very low electron density and a valley-polarized nematic dPDW for a large region [see Fig.~1(c) in the main text].

To better understand the issue, we have studied the effective $t$-$J$-$V$ model on the quasi-one-dimensional system (which is like the lattice geometry in DMRG). As shown in Fig.~\ref{SM:finite}(a) for different geometries, we indeed find a comprehensive regime of the valley-polarized sPDW phase whose size increases as the respect ratio $L_y/L_x$ decreases (while the total lattice number $L_x\times L_y$ is fixed).  
In a sense, the more the system resembles a quasi-1D geometry, the stronger the tendency for the ground state to exhibit the s-wave PDW.
Thus the above VMC results are qualitatively consistent with those obtained by DMRG (except for commensurate in VMC and weakly incommensurate in DMRG). We conjecture that whether the system is commensurate or weakly incommensurate may depend on the size and shape of the lattice under study.
To observe the incommensurate nature of the PDW order, we have performed VMC to calculate the effective $t$-$J$-$V$ model (at fixed $\omega_{\text{D}}/|t|=3$ and $n\approx 0.05$) on a lattice with a much larger system size. 
As can be seen in Fig.~\ref{SM:finite}(b), we compare optimized energies of valley-polarized sPDW states with a few nearby momentum $\pmb qs$ around the $\pmb K'$ point.
Although the energies are very close, the energy of the sPDW state with momentum $A_4$, which is away from the commensurate $\pmb K'$, is the lowest among them.
Therefore, we believe that, if a sufficiently large lattice size (say $L_x=L_y\geq 72$) can be studied by VMC, the PDW ground state should exhibit incommensurate behavior, consistent with the DMRG study.
In addition, as shown in Fig.~\ref{SM:finite}(a) for different sizes, the phase boundary between valley-polarized sPDW and dPDW has no notable changes in VMC calculations from the $18 \times 18$ to $21 \times 21$ lattice (more like 2D geometry), implying that there is a negligible finite-size effect up to $21 \times 21$.  \\

\end{widetext}

\end{document}